# Government and Social Media:
# A Case Study of 31 Informational World Cities

Agnes Mainka
Heinrich Heine
University Düsseldorf
Agnes.Mainka@hhu.de

Sarah Hartmann
Heinrich Heine
University Düsseldorf
S.Hartmann@hhu.de

Wolfgang G. Stock
Heinrich Heine
University Düsseldorf
Stock@phil.hhu.de

Isabella Peters
Heinrich Heine
University Düsseldorf
Isabella.Peters@hhu.de

**Abstract**

*Social media platforms are increasingly being used by governments to foster user interaction. Particularly in cities with enhanced ICT infrastructures (i.e., Informational World Cities) and high internet penetration rates, social media platforms are valuable tools for reaching high numbers of citizens. This empirical investigation of 31 Informational World Cities will provide an overview of social media services used for governmental purposes, of their popularity among governments, and of their usage intensity in broadcasting information online.*

## 1. Introduction

Enhanced ICT infrastructures and the increasing use of technology have reshaped communication. In 2013, YouTube and Facebook reached over one billion active users [1, 2]. These and other social media platforms have become popular in citizens' everyday lives, and municipalities, too, will use these channels in order to get in touch with citizens online. ICT infrastructure and high internet penetration are important preconditions for guaranteeing that a large amount of citizens is able to make use of those communication channels and of the internet in general.

Both these factors can be found in Informational World Cities [3, 4], meaning prototypical cities of the knowledge society (such as Singapore, New York, or Hong Kong). Informational Cities consist of two spaces: the space of places and the space of flows. The space of places (i.e., buildings, streets) is dominated by the space of flows (flows of money, power, and information). Those cities are metropolises of the 21st century. Following Manuel Castells' notion on "Informational Cities" [5], we use the concept of "Informational World Cities" [6]. A "World City" is defined by its degree of "cityness" [7, 8, 9], where a large population does not necessarily constitute an Informational World City. But infrastructures such as those in a digital city [10], ubiquitous city [11], smart city [12, 13], knowledge city [14], or creative city [15, 16] should be a given. These frameworks guarantee that a city meets the needs of the space of flows-dominated knowledge society.

Our analysis of 126 references revealed that this literature set contained advice for 31 cities with typical properties of Informational World Cities [6]. Strikingly, those cities are also global centers distributed all over the world (Figure 1). Our empirical investigation of indicators for Informational World Cities showed that the maturity of eGovernment is one crucial aspect worth intensive exploration.

In the literature, a distinction is made between the terms *eGovernance*, *eGovernment,* and *Government 2.0*. eGovernance is used as a generic term for planning, innovation, and funding at the city level [17]. According to Yigitcanlar, eGovernance is the fundamental basis for innovation in an Informational World City [18]. It comprises important properties of a city, such as the improvement of living standards and the increase of economic growth through better cooperation between authorities and citizens as well as businesses. The increasing usage of ICT allows businesses and citizens to engage in political debates and decision-making processes online [17, 19, 20]. Five interaction levels are specified by Moon to describe the function of eGovernment [21]: information, communication, transaction, integration, and participation. The second stage, "communication," has been evolving from face-to-face conversation in the office and "snail mail" correspondence to real-time conversations on social media platforms such as Facebook or Twitter [22, 23]. The fifth stage, "participation," describes an ideological notion. At this stage citizens are provided, and seize, the opportunity of engaging in political decision-making processes. The increasing governmental usage of the web is referred to as Government 2.0 [24, 25]. The term "Government 2.0" is not to be equated with "Web 2.0" [26], which was coined regarding emergent services of social media. Government 2.0, on the other hand, is used in conjunction with *"a more open, social, communicative, interactive and user-centered version of e-government"* [27]. Online interaction with citizens on governmental homepages or on social media platforms should be citizen-centered [25]. Citizens should be seen as customers whose demands need to be satisfied. Theoretical benefits of this technology include cost reductions, enhanced participation, transparency, and enhanced trust [28, 29]. Most analyses of governmental social media usage are generally performed in a more in-depth fashion, e.g. investigating communication on specific platforms and concerning specific topics [33, 34].

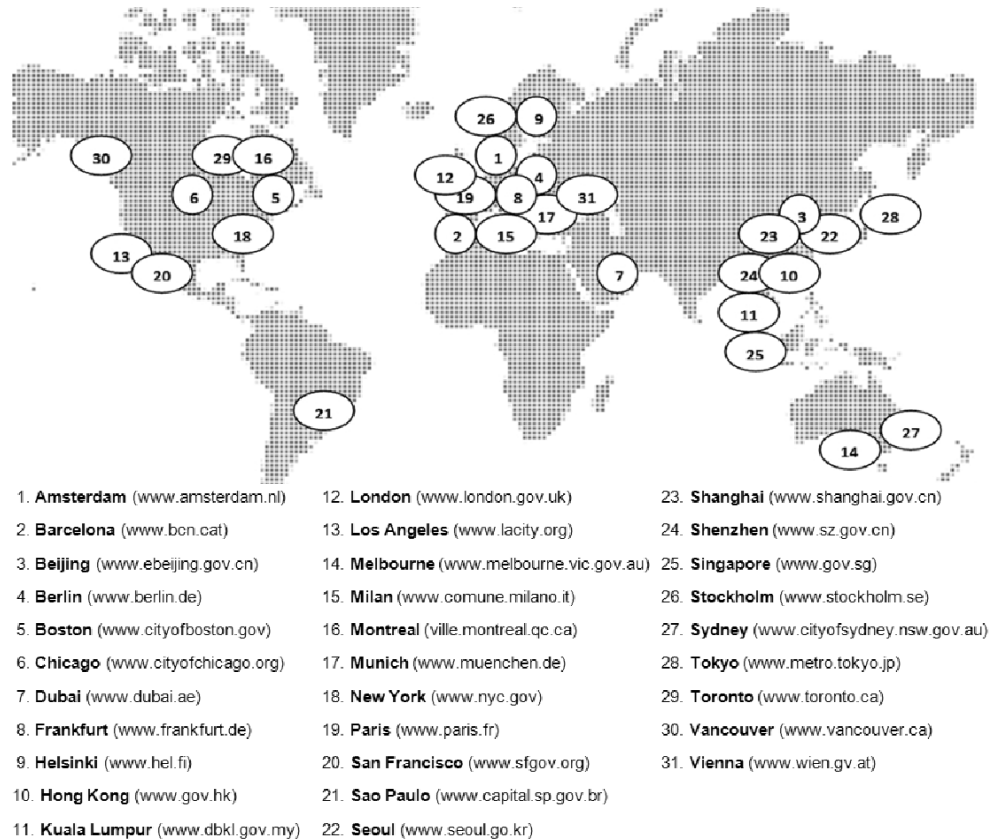

**Figure 1.** Informational world cities and the URLs of their official websites.

A study of Münchener Kreis [35] evaluated the needs of citizens in interacting with governments and administrations. This study was conducted in 2012 and 2013 and represents internet users between the ages of 18 and 70 and living in Germany, the U.S.A., Brazil, China, India, and South Korea. As shown in Table 1, more than 40 % of users in Brazil, China, and India would like to use electronic services for citizens via social media platforms. Furthermore, citizens are interested in being involved in political decision-making processes such as policy debates. However, users in all these nations have less confidence in these platforms when it comes to the protection of their personal data. This study shows that there is an audience on the world wide web who would like to use social media to get in touch with governments and administrations. Accordingly, it is advisable for governments to be represented on social media platforms if they want to reach as many of their citizens as possible.

The free market was an early adopter of social media platforms as marketing tools [30]. Governments have inherited this strategy, but social media platforms do not run themselves and being present on them does not necessarily entail eParticipation [31]. First, a strategy is needed. There are two main challenges: 1. every social media account must be continuously updated, and 2. every government must find the most profitable way of reaching its citizens [32].

When analyzing eGovernment in Informational World Cities with regard to the indicators "communication" and "participation," it must first be proven whether and which communication platforms (e.g., social media) are used by governments, whether government accounts are visible on the web, whether they are used frequently, and whether they reach a large audience. Hence, this research considers a wide range of social media platforms that are examined in light of the following four research questions:

(1) Which social media services are used by governments and when did they first open up their account?
(2) Are these social media accounts interconnected with other social media platforms and governmental web presences?
(3) Do governments frequently publish social media content?
(4) Do social media users show an interest in government accounts on social media platforms by liking or following them?

**Table 1.** Citizens' opinions on interacting with governments and administrations on social media platforms [35].

| Germany | U.S.A. | Brazil | China | India | S.Korea |
|---|---|---|---|---|---|
| I would like to use electronic services for citizens on Facebook and other social media platforms. | | | | | |
| 14 % | 22 % | 41 % | 41 % | 42 % | 31 % |
| I would like to be involved in policy decisions on Facebook and other social media platforms. | | | | | |
| 24 % | 19 % | 40 % | 37 % | 39 % | 18 % |
| I trust that my personal data will be handled responsibly on social media platforms. | | | | | |
| 9 % | 16 % | 17 % | 21 % | 37 % | 14 % |

## 2. Method

We started our study by investigating the 31 governmental websites listed in Figure 1. First they were checked for any links to official social media accounts representing the government of a city. The platforms thus identified were then browsed in order to check whether any governmental accounts of the 31 cities had not been referenced by an official website. Moreover, we looked at backlinks from all identified accounts to their official website in order to prove the services' government affiliation. In general, one can find two groups of accounts referenced by government websites: The first group consists of official government accounts or blogs for general purposes that refer to the city's government as a whole (e.g., the Facebook account "City Of New York"). The second group comprises accounts from governmental institutions, departments, or political persons (e.g., the account of the city's mayor or other politicians). This investigation focused on the first group of accounts because they seemed to be more sustainable, better maintained over time, and independent of a particular political party, mayor, or politician during any given legislative period.

To evaluate the governments' activity in social media, every account was either manually checked or accessed via provided APIs in order to collect available information concerning the accounts, the quantity of published content online, and the reactions of users, i.e. account creation date, date of first post or other activity, quantity of posts, tweets, photos, videos, pins, and comments, as well as followers and likes. Some social media platforms also allow for a stronger interaction with users, e.g. via comments on Facebook pages or retweets of governments' tweets. Since our study aimed at learning how often governments make use of which kind of social media, a deeper analysis of user interaction has been left for future research.

Because of their vast deviations in website structure and graphic characters, the Chinese websites were analyzed with the assistance of a Chinese native speaker in order to be able to reliably identify any referenced Chinese social networks. For the examination of other government websites we used the English or German version if available, or translated the website via Google Translate. The research was conducted between November 28, 2012 and January 3, 2013 and relies on the data which was available online at that time.

## 3. Results

In this section we present the results of our analysis as guided by the aforementioned research questions.

### 3.1. Governmental accounts on social media platforms

The 31 cities make use of a variety of social media services: they use social networking platforms like Facebook, Google+, and Hyves (a Dutch service); the business social networks LinkedIn and Xing (a German service); the location-based social network Foursquare; the microblogging services Twitter, Sina Weibo, the video platforms YouTube, Vimeo, Livestream, and Ustream; the photo-sharing applications Flickr and Instagram; and content-sharing services like Pinterest, Storify, Tumblr, and blogs. The total numbers of general government accounts for each social media service found on the websites are: 24 on Twitter, 21 on YouTube, 20 on Facebook, 11 accounts on Google+ and LinkedIn, ten on Instagram, seven on blogs and on Flickr, six on Pinterest, four on Foursquare and Vimeo, two on Weibo and Xing and, finally, one each on Livestream, Ustream and Tumblr. Inactive accounts (e.g., registered accounts without any posts, photos, videos etc.) were included in our analysis. Storify and Hyves have not been considered, because none of them were used by governments for general purposes. New York's blog on Tumblr is counted among blogs since it serves the same purposes.

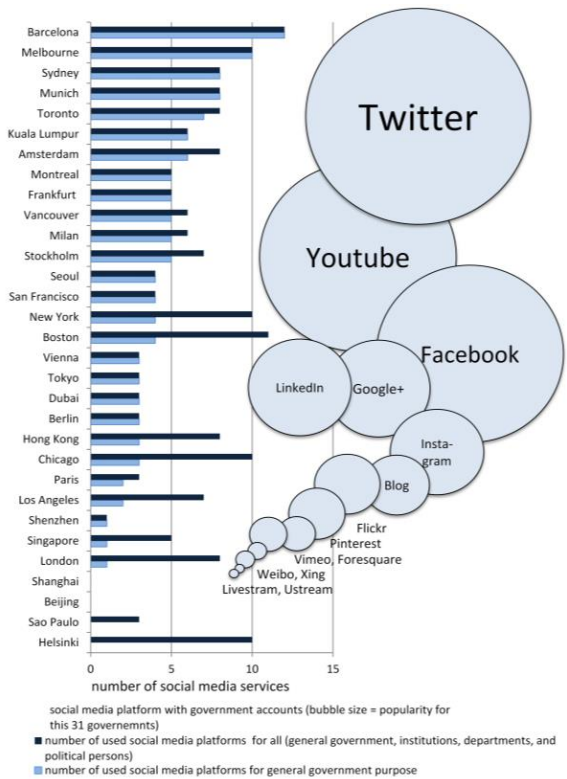

**Figure 2.** Social media platforms used for governmental purposes in Informational World Cities.

Figure 2 illustrates which social media platforms are used by governments. The services are ranked by the number of cities that use them. The most frequently used service is Twitter. All in all, 24 cities use this microblogging service for general government accounts. The Chinese microblogging service Weibo is used by only two cities (Hong Kong and Shanghai), which is due to the language barrier and political restrictions in China often keeping non-Chinese cities from using the service. Where applicable, the results of Twitter and Weibo will be cumulated due to the great resemblance between these two microblogging services. The second and third most common platforms for government accounts are YouTube and Facebook. After these, there is a fall-off in the number of cities that use a specific platform, e.g. Google+, LinkedIn, Instagram, blogs (including Tumblr), Flickr, Pinterest, etc., for general government accounts. In sum, we detected fifteen (or sixteen, including New York's Tumblr) different social media platforms used to represent general government accounts. Figure 2 also illustrates the number of social media platforms used for each city. The cities with the most diverse usage of social media services are Barcelona with twelve, Melbourne with ten, Sydney and Munich with eight and Toronto with seven general government accounts across different platforms. There are also cities that have no general government account but often use social media services to distribute information about administrations, institutions and political persons. For instance, Helsinki has no general government account but uses a very detailed social media page (Figure 3) to link to 57 different Facebook accounts, all related to Helsinki. They also refer to many accounts on Twitter and YouTube.

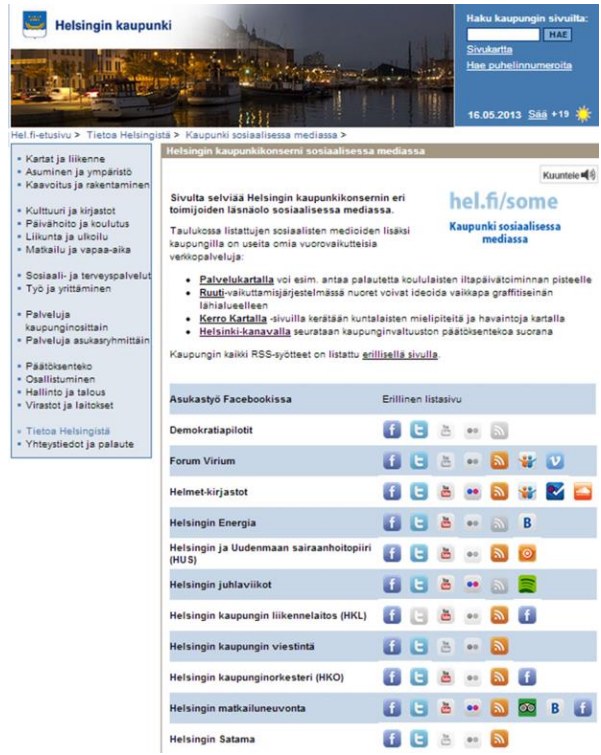

**Figure 3.** Helsinki's official government website linking to diverse social media platforms.

### 3.2. Interconnectedness between governments' social media accounts and web presences

To reach an audience online, the governments' activity on social media platforms has to be made visible to citizens and to other users. Therefore, we checked whether these accounts are linked to each city's official government website. Keeping in mind that the world wide web is considered to be a network of links [36], links from an official government website may enhance the visibility and popularity of the government accounts and the government websites, respectively. In addition, such outlinks emphasize the seriousness of government accounts on social media platforms and enhance their credibility. As shown in Table 2, most of the general government accounts link back to their official websites. Only the accounts on

Instagram, Vimeo and Ustream show few or no backlinks to their governmental parent sites. Additionally, most of these accounts on Instagram and Vimeo are inactive. It might be assumed that inactive accounts without backlinks are not official government accounts. A strong interconnectedness between the websites and social media accounts verifies, to a certain degree, the accounts' authenticity. However, a lack of backlinks to governmental websites on or inlinks from government websites to active accounts does not imply fraud, since official government accounts can also be verified via government-specific labels, designs or content. In any case, we identified inactive accounts with inlinks from governmental websites and were able to authenticate them. Accounts without any activity and links (inlinks and backlinks) could not be reliably verified as official government accounts but are considered in our data analysis.

**Table 2.** Interconnectedness between government websites and social media platforms.

| Social media platform | Outlinks from city's government website: Number of cities | Backlinks to city's government website: Number of cities |
|---|---|---|
| Twitter | 24 | 23 |
| YouTube | 21 | 21 |
| Facebook | 20 | 18 |
| LinkedIn | 11 | 10 |
| Google+ | 11 | 10 |
| Instagram | 10 | 5 |
| Flickr | 7 | 5 |
| Pinterest | 6 | 6 |
| Foursquare | 4 | 4 |
| Vimeo | 3 | 0 |
| Xing | 2 | 2 |
| Weibo | 2 | 2 |
| Ustream | 1 | 0 |
| Livestream | 1 | 1 |

Another way of drawing users' attention to the government's social media activities is cross-linking between services. With the exception of Twitter, all services support the linking from account descriptions to other services. Table 3 shows the number of cities that outlink from one of their accounts to another social media presence and the number of cities that have an inlink from another used service on their account. We found out that there are not that many links between the governments' social media services, which might be due to the services' limited linking options. For example, 13 cities link from another service to Twitter but just two of these accounts then established links to other services, probably because of Twitter's space limitations on account descriptions.

**Table 3.** Interconnectedness between governments' social media accounts.

| Social media platform | Outlinks to government account: Number of cities | Inlinks from government account: Number of cities |
|---|---|---|
| Twitter | 2 | 13 |
| Facebook | 10 | 9 |
| YouTube | 7 | 6 |
| Pinterest | 6 | 5 |
| Blogs | 4 | 3 |
| Foursquare | 5 | 3 |
| Google+ | 1 | 3 |
| Flickr | 0 | 3 |
| Livestream | 1 | 2 |
| Instagram | 1 | 2 |
| Ustream | 1 | 1 |

### 3.3. Social media activity

The third research question is dedicated to the activity of government accounts on social media platforms. This activity was measured via the amount of posted content on each service. Interactive activities, such as comments from users and retweets on Twitter, were not studied here as we focus on the governments' activity in our evaluation of the extent to which they use social media and of how much content they produce overall. The results can then be used as starting points for further studies examining the reactions of users and their degree of engagement with governments' accounts and content. However, the amount of posted content depends, for one, on the time span during which a service has been used and on the effort that has been made to create certain contents. Therefore, we calculated the average quantities of posted content per month, across all cities, for the following platforms: Twitter, Flickr, blogs, Instagram, Foursquare, and YouTube. Weibo and Pinterest had to be excluded since they do not provide account creation dates.

As shown in Figure 4, the highest rate of activity was found on Twitter with 135 tweets per month and per city using Twitter. Flickr, with 39 pictures per month and city, is also used intensively and is more popular than Instagram with its five pictures. Surprisingly, blog posts (22 posts) are also very popular even though they take longer to produce than pictures or tweets. Each month the 31 government accounts produce five tips on Foursquare and five videos on YouTube. The latter is often used to broadcast local events, news, and reports. On Foursquare, the number of references to other locations or short posts was shown.

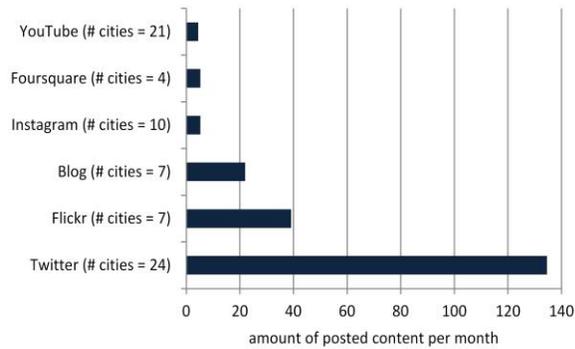

**Figure 4.** Average activity of government accounts on social media platforms across cities per month.

Figure 5 displays the amount of monthly published content on each social media platform for each city. The results show that there is a significant difference between the governments. Beijing, Shanghai, Helsinki and Sao Paulo contribute no content at all, and are thus excluded from our analysis. In contrast, the government accounts of Berlin, Seoul, and Barcelona publish more than 500 posts per month. However, Berlin only publishes content on Twitter (563 tweets per month) whereas Seoul's strategy is focused on the two platforms Twitter and blogs, with nearly 500 tweets per month and 40 blog posts. Barcelona is one of the few cities using more than three services and is very active on Twitter (more than 300 tweets per month), Flickr (70 pictures per month), and Instagram (nearly 30 pictures per month). It is also represented on YouTube, Foursquare and blogs. Most of the cities use two services at most. Of crucial importance to our evaluation of the amount of content published by cities is their respective period of participation; hence, we also examined when the governments first started their activities in social media relative to the average starting time of all analyzed cities. As illustrated in Table 4, Sydney was the first city to register any social media accounts, i.e. official general government accounts on Flickr and YouTube. Flickr, Twitter, and YouTube are the longest-used social media services over any average period of all government accounts. Stockholm seems to have been active in social media for a long time as well. It was the first city to run a blog and a microblog. The first Facebook page was created by San Francisco in November 2008.

There are long time spans between the launches of social media services and their factual use by governments, e.g. Facebook was launched in 2004 but the average join date for governments was six years later, in 2010. This may be due to the typically delayed uptake of social media activities by businesses and governments. Not before 2009 did social media start to truly establish themselves in the business world [37] and in governments, many of which built up a systematic presence [38]. In contrast, services launched at a later date, such as Google+, Instagram and Foursquare, were quickly adopted by governments.

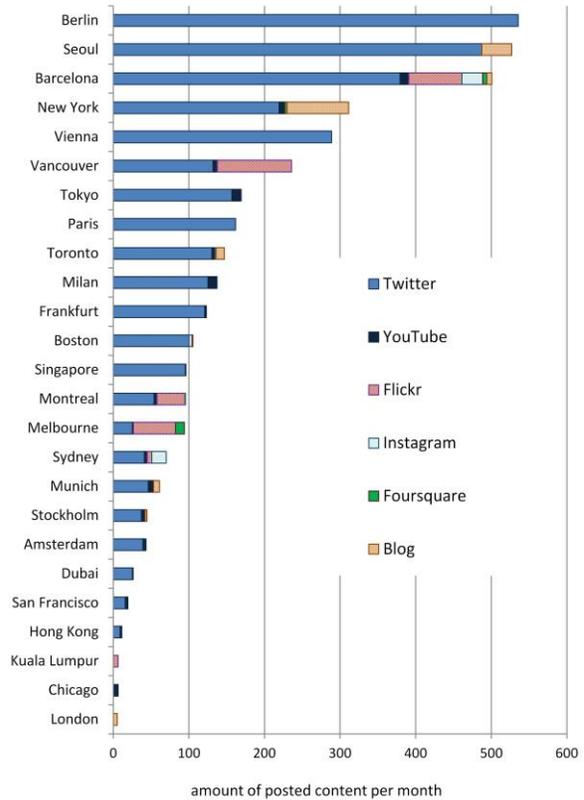

**Figure 5.** Cities' government activity on social media platforms per month.

The same behavior can be shown for Flickr or YouTube, where the time difference between the first cities' join dates and the average join date for all cities is more than three years. Twitter became a popular service among most cities almost at the same time as Flickr and YouTube, although Twitter's first government account was registered in August 2008. Cities have been actively blogging since January 2008, far earlier than they took up Facebook and Twitter. Less extensive and more recent usage numbers are available for Ustream, Google+, Instagram, and Foursquare. Account creation dates for Weibo, Livestream, Pinterest, LinkedIn and Xing are missing because those services do not publicly provide this information.

### 3.4. Followers and likes on social media platforms

Our last research question regards the success of government accounts on social media platforms, and will serve to clarify whether government accounts adequately address social media users. Therefore, all accounts were examined with regard to their number of followers, likes, subscribers, etc., which can be compared to the number of unique visitors on a website with an additional news subscription. Almost all social media services provide information about their number of user subscriptions. Blogs are excluded, since they do not aim at forming a network of people. Subscriptions to blogs via RSS-Feeds are possible, but are not made visible on the blog itself.

As displayed in Figure 6, the platform with the most user subscriptions is Facebook, with more than five million likes across cities. On average, there are about 154,000 likes for each city using Facebook. The collective number of Twitter and Weibo followers is, at more than 1.3 million, also very high. The average value is about 55,000 followers per city using Twitter. Fewer subscribers are reached on Foursquare, where the average value is at slightly more than 10,000 likes, and on LinkedIn and Xing, where an average of 3,400 contacts was calculated. At about 1,300 users in the circles of government accounts, Google+ has fewer subscribers. YouTube and Instagram, with about 500, and Pinterest, with 200 subscribers on average, are of marginal importance in this area.

**Table 4.** When did analyzed cities create an account on social media platforms? Comparison between the earliest and the average join dates of governments.

| Social media services | First government account online | Governments average entry date | Social media services' launch |
|---|---|---|---|
| Facebook | San Francisco in 11/2009 | 09/2010 | 02/2004 |
| Google+ | Melbourne in 11/2011 | 12/2011 | 06/2011 |
| Twitter | Stockholm in 08/2008 | 11/2009 | 07/2006 |
| YouTube | Sydney in 10/2006 | 11/2009 | 02/2005 |
| Ustream | Seoul in 04/2011 | 04/2011 | 03/2007 |
| Flickr | Sydney in 08/2006 | 11/2009 | 02/2004 |
| Instagram | Toronto in 06/2011 | 01/2012 | 10/2010 |
| Foursquare | Barcelona in 08/2011 | 03/2012 | 03/2009 |
| Blogs | Stockholm in 01/2008 | 05/2010 | Since 1990 |

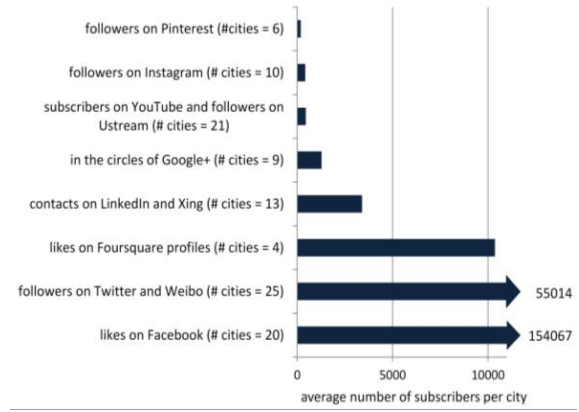

**Figure 6.** Average number of subscribers per city on social media platforms.

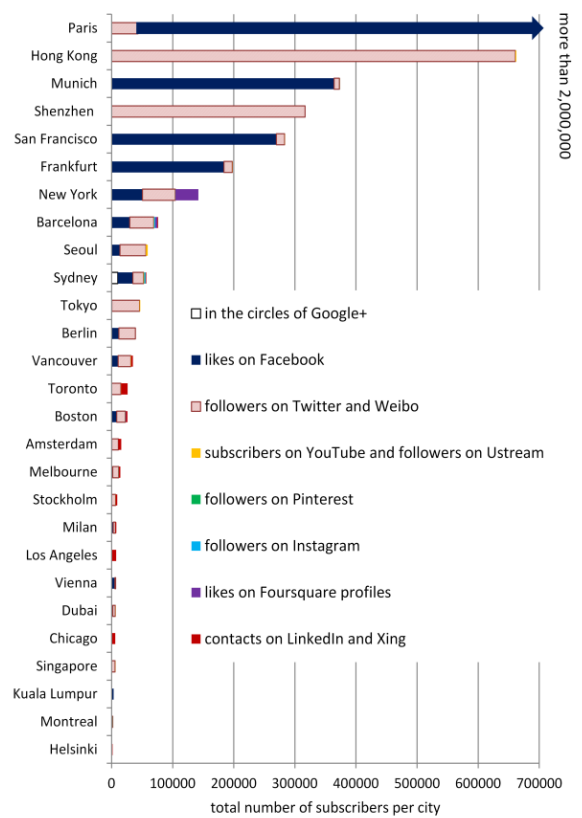

**Figure 7.** City-wise comparison of governments' number of subscribers on social media platforms.

Figure 7 presents the summarized numbers of followers and likes for general government accounts on social media platforms per city. It is conspicuous that, at more than two million likes, Paris attracts far more Facebook likes than any other analyzed city. The government accounts on Facebook for Munich, San Francisco, and Frankfurt all reach more than 180,000 likes as well. Although their social media activity is limited to Weibo, Hong Kong and Shenzhen reach about 600,000 and 300,000 followers, respectively. In general, the numbers of followers and likes differ greatly between the analyzed cities. Some cities, i.e., Paris, Hong Kong, Munich, Shenzhen, San Francisco, Frankfurt, and New York, are very good at collecting subscribers online. The other cities only have very few or even minimal numbers of followers.

As has been observed in the case of activity numbers, users of government social media accounts concentrate on two or three services at the most when following or liking cities. Twitter and Facebook are the most-subscribed services by users, and also the most used services by governments. Conspicuously, YouTube, which is the third service with a high number of government accounts, only has a low number of subscribers. By contrast, Foursquare is only used by four cities for general government accounts but each activates more users than many accounts on other services do.

## 4. Discussion

Our analysis of government activities across social media was conducted for 31 Informational World Cities. The assumption was that cities which are important metropolises in the 21st century use social media services for government-to-citizen interaction. Our results show that there are strong differences between the respective popularity of social media services for each city. The most-used social media platforms are Twitter, YouTube and Facebook. All in all, 15 (with Tumblr, 16) different social media platforms are used by the 31 cities. Twenty-nine cities use at least one of the social media services; and on average, four services are used per government.

Weibo is the most popular social network in Chinese regions [39], providing Twitter-like functionalities. Hong Kong and Shenzhen reach very high numbers of followers, placing second and third in this study in terms of subscribers, respectively. The two other Chinese cities, Shanghai and Beijing, do not use any social media services. This could be due to the access restrictions to globally accessible social media services imposed by China's government. In contrast, Beijing and Shanghai have their own government microblogs under construction, presumably to exercise better content control with regard to their political practices. Hence, the comparability of results between Chinese cities and the others is not given.

It is remarkable that many cities with government accounts on social media platforms do not link to them from their homepage. Some cities, like Helsinki, subscribe to another information policy and have a special webpage (Figure 3) where all social media activities are listed, but some do not link to any account on Facebook or Twitter from their homepage. It can probably be assumed that a lack of links from the government's homepage to its social media services hamper citizens' participation, which results in lower numbers for followers and likes (except for Hong Kong). In this study, Twitter is not only the most popular social media application (in terms of governments posting content) but also the service with the highest amount of activity (in terms of users liking content or following accounts). This is not surprising as microblogging only takes a few seconds, in contrast to creating video clips for YouTube, which requires a greater effort. Therefore, the services' functionalities and differences in terms of usage cause different user behavior and should be kept in mind when comparing user statistics. Interaction numbers (i.e. comments) on Facebook were not considered in this analysis because of the lack of information on Facebook profile pages. In terms of followers and likes on social media platforms, Facebook is the most effective service in terms of animating users to like government profile pages, compared, for example, to Google+. However, Google+ is a very young service compared to Facebook or Instagram. Thus it is not only the differences within the services that must be considered, but the services' periods of activity are equally important for deciding whether they are appropriate for government-to-citizen communication.

Whether or not governments do, in fact, reach their citizens cannot be answered by this study. However, an attempt to answer this question can be made if we assume that cities with a high population are more likely to obtain increased numbers of city-based Facebook users and likes. In both cases, a strong positive correlation between both values is to be expected. We found that the Pearson coefficient between the number of a city's inhabitants and the number of city-based Facebook users is $r = +0.87$, but $r$ is -0.22 when correlating the number of city-based Facebook users with the number of likes. The latter indicates that it is not necessarily the Facebook users based in the particular city who are responsible for the popularity of the Facebook page. Unfortunately, no other social media services provide the number of users per city, meaning that the city-wise correlation cannot be investigated further. When using the available numbers of a city's general population and the number of Facebook likes for government accounts, the Pearson correlation (two-sided) arrives at $r = -0.26$. In contrast, calculations with Twitter followers ($r = +0.42$) as well as YouTube subscribers ($r = +0.51$) and the number of the city's inhabitants show a stronger positive correlation. For the other services, only a low correlation was found. Due to the lack of city-specific user data, we used general population numbers for calculating the correlations between numbers of likes, followers or subscribers of a particular account. The results might be misleading, however, since social media services are available to every internet user and not only to those based in the respective city.

To put it in a nutshell, our study showed that no Informational World City is more prominently active than all the others. Nevertheless, there are tendencies for each city to be more or less active in social media services. In general, more activity engages more users, but there are a lot of factors that can affect the numbers of followers and likes, as shown by the follower numbers of Paris, which outclass all other cities. However, the cities' popularity and population size must also be regarded, e.g., Paris might be more popular than Helsinki in terms of tourism or events.

## 5. Conclusion and Future Work

Twitter, YouTube, and Facebook are the social media services most used by governments. Which services are most frequently used by users is difficult to examine due to several reasons. First, there is a lack of data concerning the services, which is why not all services could be compared. Second, there are differences in the services themselves, which results in divergent user behavior (e.g., concepts of likes and followers), and third, there are differences in the time spans of activity, e.g., upcoming services do not have numbers as high as those of established ones but can become very popular quickly. Concerning activity, Twitter is the service with the highest number of posts by far, and in terms of followers and likes, Facebook and Twitter are of capital importance [40]. YouTube is conspicuous in that almost all governments are present here but their accounts are less often subscribed to by their users. YouTube did not achieve high numbers for either activity or subscribers although it is one of the top three services used by governments in Informational World Cities. Nevertheless, governments in Informational World Cities do reach users with their social media activities, provided they choose the most appropriate services for their government-to-user communication. However, only a few services achieve high numbers of users. Accordingly, we may conclude that only two or three services are sufficiently capable of reaching citizens. Furthermore, lower usage numbers do not have to be due to low user participation. Presumably, many users watch YouTube videos without subscribing to the respective YouTube channel. To summarize, governments should keep an eye on upcoming services and use those that their citizens also use. Additionally, cultural differences must be considered, e.g., Twitter does not work in Chinese regions.

Another important point for analyzing social media usage is the content of the actual accounts (e.g., what type of information is provided by the governments), the types of posts (e.g., are there text posts only or videos etc. as well?), and the user-created content (e.g., what do users post on government accounts). Therefore, a more detailed content analysis as well as a more differentiated analysis of users must be conducted in the future. Do governments reach the "real" citizens and do they try to provoke comments and discussions about future visions of the city? It must be tracked whether these discussions are officially considered in the cities' governance or if social media services are only used to broadcast news.

## Acknowledgment

We thank Paul Becker for checking spelling and grammar of our text. The work is funded by the Strategic Research Fund of the HHU Düsseldorf, Germany (F 2012 279-10).